\documentstyle[aps]{revtex}
\newcommand{\be}{\begin{equation}}
\newcommand{\ee}{\end{equation}}
\newcommand{\bea}{\begin{eqnarray}}
\newcommand{\eea}{\end{eqnarray}}
\newcommand{\FN}[1]{$\typeout{Kommentar: #1}$}

\begin{document}               
\bibliographystyle{prsty}


\title{The Stability of Standing Waves with Small Group Velocity}
\author{Hermann Riecke}
\address{Department of Engineering Sciences and Applied Mathematics\\
Northwestern University, Evanston, IL 60208, USA} 
\author{Lorenz Kramer}
\address{
Physikalisches Institut, Universit\"at Bayreuth, D-95440 Bayreuth, Germany}
\maketitle

\begin{abstract}

We determine the modulational stability of standing waves with small group velocity
in quasi-onedimensional systems
slightly above the threshold of a supercritical Hopf bifurcation. 
The stability limits are given by two different long-wavelength destabilisation 
mechanisms and generically also a short-wavelength 
destabilisation. The Eckhaus parabola is shifted off-center and can be convex from 
below or above. 
For nonzero group velocity the Newell criterion, which near the cross-over from standing to 
traveling waves becomes a rather weak condition, does not determine the destabilisation
of all standing waves in one dimension. The cross-over to the non-local
equations that are asymptotically valid near threshold is discussed in detail.
Close to the transition from standing to traveling waves complex dynamics can arise
due to the competition of counter-propagating waves and the wavenumber selection 
by sources.
Our results yield necessary conditions for the stability of traveling 
rectangles in quasi-twodimensional systems with axial anisotropy and
form a starting point for understanding the spatio-temporal chaos of traveling oblique rolls 
observed in electroconvection of nematic liquid crystals. 


\end{abstract}
\ \\

\section{Introduction}

In this work the coupled complex Ginzburg-Landau equations  
\bea \label{e:CCGLE}
\partial_t A+ u \partial_y A = \mu A + ( 1+ ib_2) \partial_y^2 A 
 -c|A|^2A - h|B|^2A,\label{e:swA} \\
\partial_t B- u \partial_y B = \mu B + ( 1+ ib_2) \partial_y^2 B
 -c|B|^2B - h|A|^2B,\label{e:swB}
\eea
describing the interaction of two counterpropagating waves arising via a 
supercritical Hopf bifurcation in one space dimension (see e.g. \cite{CrHo93}) are 
investigated. 
Here $A$ and $B$ are the complex wave amplitudes (or envelopes).
The coefficients $c$ and $h$ are complex (we write $c=c_r+ic_i$ etc.) with $c_r>0$
and $c_r+h_r>0$ (see below), all other coefficients are real. 
The (dimensionless) control parameter $\mu$ ($>0$ above threshold) carries the system 
across the instability.
Time and length are scaled such that the diffusion constant is $1$ \cite{scale}.
Uniform scaling of Eqs.(\ref{e:swA},\ref{e:swB}) with $\mu^{3/2}$ will be obtained by introducing 
slow variables $T=\mu t$ and $Y=\mu^{1/2} y$, and scaling $A,B$, and the group velocity $u$ 
with $\mu^{1/2}$. 
Strictly speaking this requires $u$ to be small of order $O(\mu^{1/2})$. 
While we focus on this case, 
we also discuss the case $u=O(1)$ as well as the cross-over between the two cases.
We expect that our results will give 
insight into the case $u=O(1)$ somewhat further above threshold.

The spatially varying part of the physical fields ${\bf u}$ are given in terms of these 
variables as
\be
{\bf u}=\mu^{1/2}  e^{i \hat{\omega}_h \hat{t}}\left(A(y,t) e^{i(\hat{p}_c \hat{y})} {\bf f}_1(x,z) +
B(y,t) e^{i(- \hat{p}_c \hat{y})} {\bf f}_2(x,z) \right)+ O(\mu) + c.c.\label{e:defv}
\ee
where $\hat{\omega_h}$ is the frequency of the fastest-growing 
linear mode, which at threshold ($\mu = 0$) is the Hopf frequency,
$\hat{p}_c$ the critical wavenumber and 
${\bf f}_{1,2}(x,z)$ 
eigenfunctions of the linear problem. The hatted quantities (e.g. $\hat{t}$)
are unscaled physical quantities.

In the case $h_r>c_r$ 
the two waves suppress each other and one often needs to consider
only a single equation describing a traveling wave train. 
In many cases the group velocity term can then be discarded by going into a moving frame. 
The remaining equation has been studied intensely.
In particular it is well known that the plane-wave solutions with wave number $p_0$
are stable with respect to long-wave side-band perturbations inside the band restricted by
\be \label{e:travstab}
1+\frac{b_2 c_i}{c_r} - 2 \frac{\left(1 + (c_i/c_r)^2\right) p_0^2 }{c_r F_{TW}^2} > 0,
\ee
where $F_{TW} = \sqrt{\mu-p_0^2}/c_r$ is the amplitude of the wave.
Thus the band center solution $p_0=0$ is the last to lose stability when the Newell
criterion $c_r+b_2 c_i > 0$ becomes violated \cite{Ne74,StPr78}. In addition, for 
strong dispersion short-wave side-band perturbations can be important \cite{StPr78,MaVo93}. 
This occurs only for wavenumbers away from the band center 
and when the Newell criterion is satisfied ($c_r+b_2 c_i > 0$).  It 
is not relevant in the following.

The coupled equations become important when $h_r^2$ becomes smaller than (or only slightly
larger than) $c_r^2$ and then the superposition of oppositely traveling waves,
leading to standing-wave solutions, becomes important. 
The stability of standing waves with respect to
side-band perturbations has been studied so far only in a few special cases 
\cite{Mi95,CoFa85,Sa95,Kn95,MaVo93}. 
The case $u=0$ and $h_i/h_r=c_i/c_r$ has been investigated in quite some detail 
in the context of polarized lasers \cite{Mi95}. 
There, however, the effects discussed below do not arise.
For general values of the 
coefficients the solution in the center of the wave-number band has been
investigated, for which the stability analysis is substantially simplified
\cite{CoFa85,Sa95}. In addition to a long-wave instability an instability arising first
at a finite perturbation wavenumber has been identified. The nonlinear evolution
of the long-wave instability has been studied using coupled phase equations \cite{CoFa85,Sa95}.
The evolution arising from the instability at finite wavenumber has been studied
through numerical simulations of the Ginzburg-Landau equations \cite{Sa95}.
The behavior away from the band center has not been studied in detail.
From previous work we know, however, that 
the instability of the band-center solution may not always reflect the behavior of
the solutions with other wave numbers \cite{Ri90a}. 
We therefore present here a comprehensive analysis 
of standing waves for the complete band of wave numbers.

We know of no physical (quasi-) onedimensional
 system in which standing waves have been
 clearly observed. \FN{(nonlinear {optics???})}
In thermal convection of binary mixtures in porous media in the Hopf bifurcation
range rather strong oscillations of the Nusselt number have been observed \cite{ReAh86}, 
which can be taken as a hint for their occurrence in that system.
Theoretically, a Hopf bifurcation  to standing waves has been predicted 
for rotating convection at small Prandtl numbers  \cite{ClKn93}. 

\begin{figure}[htb]
\begin{picture}(120,240)(0,0)
\put(120,0) {\includegraphics{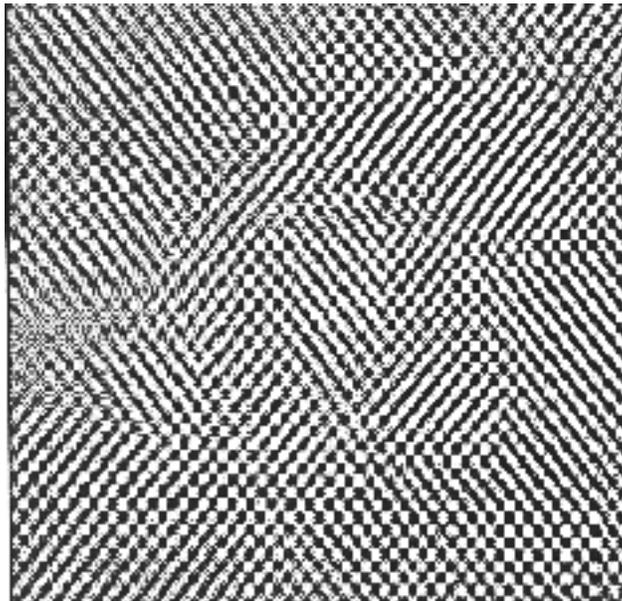}}
\end{picture}
\caption{Oblique waves (`zigs' and `zags')
 and traveling rectangle patterns observed experimentally in electro-convection 
of nematic liquid crystals \protect{\cite{DeCa95}}. \protect{\\}
\protect{\label{f:ehc}}
}
\end{figure}

In quasi-twodimensional systems with axial anisotropy an 
analogous situation arises when 
oblique rolls, i.e rolls appearing at an oblique angle with respect to the preferred axis,
superpose to give traveling rectangles.
In fact
our investigation is motivated in particular by recent experiments on electroconvection in 
thin layers of nematic liquid crystals in the usual geometry with planarly aligned director, 
(see e.g. \cite{KrPe96}) where 
in some parameter range Dennin et al. \cite{DeCa95,DeAh96,DeCa98}
 found near onset 
extended small-amplitude spatio-temporal chaos characterized by the interaction of 
obliquely traveling waves (rolls) and defects in these waves.
Figure \ref{f:ehc} shows a typical state \cite{DeCa95}.
It  consists of patches of waves traveling obliquely to the direction singled out by the 
anisotropy of the liquid crystal and their superposition, which can be considered as 
traveling rectangles \cite{SiRi92}. A starting point for the understanding of this state 
of spatio-temporal chaos is therefore the consideration of the stability
of traveling rectangles.

The liquid crystal system is axially anisotropic and the complex dynamics arises immediately above the 
onset of convection in a supercritical bifurcation. 
Combined with the fact that the origin of the traveling waves in the liquid-crystal system
is finally understood   (on the basis of a weak-electrolyte model 
\cite{TrKr95,DeTr96,TrEb97,TrKr97})
these features promise that the chaotic state may be understood {\it quantitatively} using equations
which to a large extent can be dealt with analytically, i.e. Ginzburg-Landau equations.

A full, small-amplitude description of this system 
requires 4 complex modes for the left- and right-traveling waves in the 2 oblique directions. 
If one considers the common situation where oppositely 
traveling waves suppress each
other strongly one is lead to two coupled Ginzburg-Landau equations,
\bea
\partial_t A+ u \partial_y A &=& \mu A + (1+ib_1) \partial_x^2 A + ( 1+ ib_2) \partial_y^2 A 
+ 2a\partial_{xy}^2A -c|A|^2A - h|B|^2A,\label{e:cglA} \\
\partial_t B- u \partial_y B &=& \mu B + (1+ib_1) \partial_x^2 B + ( 1+ ib_2) \partial_y^2 B
- 2a\partial_{xy}^2B -c|B|^2B - h|A|^2B. \label{e:cglB}
\eea
In a restriction to  modulations  in the $y$-direction 
 (\ref{e:swA},\ref{e:swB}) are recovered.
In this paper we therefore capture a subspace of the possible 
instabilities of traveling rectangles. 
For a quantitative comparison with the experiments the full twodimensional equations 
using parameters determined from the hydrodyamic equations \cite{TrKr97}
will be needed \cite{TrRi98}.
Presumably the  onedimensional situation is also accessible 
experimentally by studying properly dimensioned narrow channels.

In Sec. \ref{s:sw} we study the stability of standing-wave solutions of Eq. (\ref{e:swA},\ref{e:swB}). 
We find that, surprisingly, their
stability is limited by two types of long-wave instabilities and a short-wave instability.
One of the long-wave instabilities  is related to the Eckhaus instability. Its parabolic
stability limit is, however, shifted with respect to the neutral curve and can even 
be convex from above instead of below (see fig.\ref{f:oeps} below). 
The shift results from the linear group velocity term.
Moreover, the stable solutions can be outside the Eckhaus parabola instead of inside.
Not too far from  the point of degeneracy between standing and traveling waves the Newell-type
criterion for the instability of the band-center can be satisfied even for 
weak dispersion, but it does not indicate the loss of stability of standing waves at 
all wave numbers.

Thus the group velocity $u$ plays an important role. 
In the general case of $u=O(1)$ the asymptotic validity of Eqs.(\ref{e:swA},\ref{e:swB})
as $\mu \rightarrow 0$ has been questioned and in that limit equations have 
been proposed which are coupled through a term in which the counterpropagating wave
is averaged over \cite{KnDe90,Kn92,MaVo92,Ve93}. 
The long-wave stability properties of standing waves in these equations are found to 
differ markedly from those obtained from Eqs.(\ref{e:swA},\ref{e:swB}).
This has been demonstrated explicitly for the waves 
in the band center \cite{Kn92}. 
In Sec. \ref{s:knob} we show that the long-wave stability results of 
the non-local theory can 
be recovered from those of Eqs.(\ref{e:swA},\ref{e:swB}) by taking the limit 
$\mu \rightarrow 0$ 
before allowing the wavenumber $p$ of modulations to become small \cite{Kn92}. 
This corresponds to the situation in a finite system.
The usual long-wave stability theory of Eqs.(\ref{e:swA},\ref{e:swB}), on the other hand, amounts to 
$p \rightarrow 0$ in the infinite system. Strikingly, the cross-over from the result
of the non-local equations to that of (\ref{e:swA},\ref{e:swB}) can occur 
 already very close to threshold.

\section{The Stability of Standing Waves}
\label{s:sw}
The (unmodulated) standing-wave solutions of Eqs.(\ref{e:swA},\ref{e:swB}) are given by 
\be
A=F e^{i (\omega t + p_0 y)}, \qquad B=F e^{i (\omega t - p_0 y)}\label{e:deftre}
\ee
with 
\be
F^2=(\mu-p_0^2)/(c_r+h_r),\ \omega=-u p_0-b_2 p_0^2-(c_i+h_i) F^2, \label{e:defF}
\ee
Clearly one needs $c_r+h_r>0$ in order that the standing waves bifurcate supercritically. 
The neutral curve is given by $F^2=0$.
Constant phases can be added to $A$ and $B$ separately.

The stability analysis with respect to side-band perturbations is based on the 
expansion
\bea
A=\left(1+v_1 e^{\lambda t + ipy}+ v_2 e^{\lambda^* t - ipy} \right) 
F e^{i(\omega t+p_0y)},\label{e:expA}\\
B=\left(1+w_1 e^{\lambda t + ipy}+ w_2 e^{\lambda^* t - ipy} \right) 
F e^{i(\omega t-p_0y)}.\label{e:expB}
\eea
Inserting (\ref{e:expA},\ref{e:expB}) into (\ref{e:swA},\ref{e:swB}) yields after
linearization in $v_i$ and $w_i$ a dispersion relation 
in the form of a quartic polynomial for $\lambda=\lambda(p)$ for 
given wave number $p_0$ 
and control parameter $\mu$. Alternatively, through (\ref{e:defF}), 
the amplitude $F$ can be considered as the control parameter. 
First of all, for $p=0$, one finds a doubly degenerated eigenvalue $\lambda=0$ 
corresponding to the two phase modes, and two amplitude modes with 
$\lambda=-2(c_r\pm h_r)F^2$. Thus the conditions $c_r>0$ and $c_r^2>h_r^2$ are needed for 
amplitude stability. 
The long-wave expansion of the dispersion relation of the relevant phase mode
\be
p=\epsilon p_1,\qquad \lambda = \epsilon \lambda_1 + \epsilon^2 \lambda_2 + ...,\qquad 
\mbox{ with } \epsilon \ll 1,\label{e:defla}
\ee
can be treated analytically. We find non-trivial instabilities at orders $O(\epsilon)$ as well as  
$O(\epsilon^2)$. Our analysis of those instabilities shows that quite generally one has to consider 
also instabilities at finite values of $p$. 
This is in contrast to the stability of traveling waves ($A=0$ or $B=0$) for which no instability 
occurs at order $O(\epsilon)$ and only in extreme cases a short-wavelength instability arises \cite{StPr78,MaVo93}. 
We have therefore also performed a general stability analysis by solving the quartic polynomial for $\lambda$ numerically. It shows that despite the importance of short-wave
instabilities the long-wave analysis provides over a wide range of parameters a framework 
for the overall structure of the stability regions.

\begin{figure}[htb]
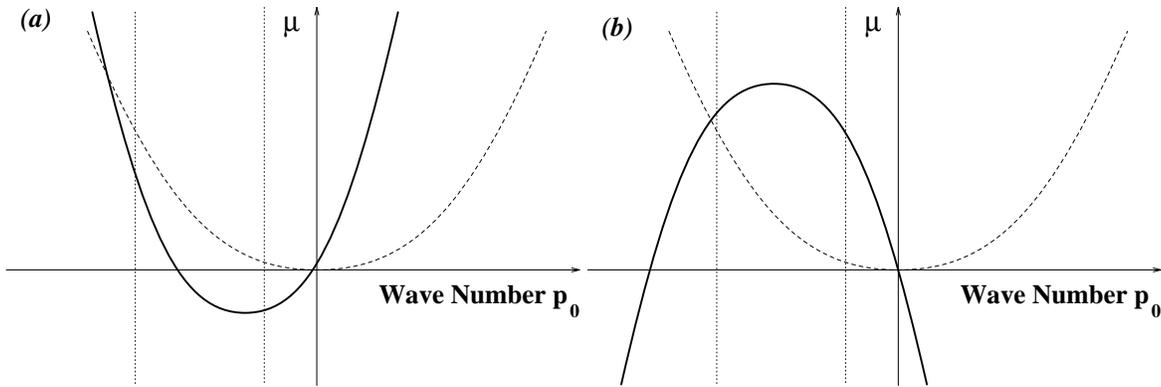

\begin{picture}(120,180)(0,0)
\put(210,0) {\includegraphics{bilder/stabisketch.2.eps}}
\put(-10,0) {\includegraphics{bilder/stabisketch.1.eps}}
\end{picture}
\caption{Sketch of possible long-wave stability limits for standing waves.
Standing waves are stable inside or outside the area delimited by the
vertical dotted lines (instability at $O(\epsilon)$), depending on whether they exclude or 
include the band center, and 
inside or outside the parabolic area given by the heavy line ($O(\epsilon^2))$. 
The dashed line denotes the neutral curve.
\protect{\label{f:oeps}}
}
\end{figure}

In the long-wave expansion one finds at leading order in $\epsilon$
\be
\lambda_1^2=-p_1^2\frac{\alpha_1 \alpha_2}{c_r^2-h_r^2}\label{e:oeps}
\ee
with
\be
\alpha_{1,2}=u(c_r \pm h_r)+2p_0\{b_2(c_r \pm h_r)-(c_i \pm h_i)\}.
\ee
Thus, to this order the solution in the band center, $p_0=0$, is always 
stable 
since $\lambda_1$ is purely imaginary. In previous work \cite{CoFa85,Sa95} therefore
no instability was found at this order. Such an instability arises, however, for $p_0 \ne 0$ (and $u \ne 0$)
when the factors in the numerator of Eq.(\ref{e:oeps}) change sign. As in a conservative system,
 the two imaginary eigenvalues then turn
into two real eigenvalues, one positive and one negative, and lead to a steady instability.
This occurs at
\be
p^{(1)}_{0,1}=-\frac{u}{2}\frac{c_r-h_r}{b_2(c_r-h_r)+h_i-c_i},\qquad
p^{(1)}_{0,2}=-\frac{u}{2}\frac{c_r+h_r}{b_2(c_r+h_r)-h_i-c_i}.\label{e:plin}
\ee
 Note, that to this order the
stability limit is independent of $F$ and therefore also of $\mu$. 
From (\ref{e:oeps}) one finds that the standing waves are unstable outside the interval 
$(p^{(1)}_{0,1},p^{(1)}_{0,2})$ and stable (to this order) inside that interval for
$\gamma_1<0$ and vice versa, where 
\be
\gamma_1 \equiv (c_i-b_2 c_r)^2-(h_i-b_2 h_r)^2
=u^2\frac{c_r^2-h_r^2}{4 p^{(1)}_{0,1}\, p^{(1)}_{0,2}}. \label{e:plininf}
\ee
Clearly this is equivalent to $p^{(1)}_{0,1}$ and $p^{(1)}_{0,2}$ having opposite or same sign, respectively.
The stability limits (\ref{e:plin}) are sketched by dotted lines in fig.\ref{f:oeps}.

At $O(\epsilon^2)$ one finds that either $\lambda_1=0$ or
\be
\lambda_2=p_1^2 \left\{\beta_0-\beta_1\frac{p_0u}{F^2}
 - \beta_2\frac{p_0^2}{F^2}\right\}\label{e:lambda2}
\ee
with
\bea
\beta_0 &=& \frac{b_2(h_rh_i-c_rc_i)}{c_r^2-h_r^2}-1,\nonumber \\
\beta_1 &=& \frac{2h_r(c_ih_r-c_rh_i)}{(c_r^2-h_r^2)^2},\nonumber \\
\beta_2 &=& 2\frac{2b_2h_r(c_ih_r-c_rh_i)-c_r(c_r^2-h_r^2+c_i^2-h_i^2)}{(c_r^2-h_r^2)^2},
\label{e:alpha13}
\eea
and $F^2$ given by (\ref{e:defF}). 
The limit $p_0 \rightarrow 0$ ($\mu \ne 0$) agrees with previous results \cite{CoFa85,Sa95}.
The Eckhaus-stability limit $\lambda_2=0$ is given 
by $\mu=\mu^{(eck)}(p_0)$ with
\be
\mu^{(eck)} = \frac{\beta_1}{\beta_0}(c_r+h_r)up_0+ K^{(eck)} p_0^2 \qquad \mbox{where} \qquad 
K^{(eck)} = 1+\frac{\beta_2}{\beta_0}(c_r+h_r). \label{e:eckhaus}
\ee
In the parameter regime in which it is relevant ($\lambda_1^2<0$) the Eckhaus instability
 corresponds to an oscillatory 
instability and issues regarding its convective and absolute character arise. 

The stability limits
given by (\ref{e:lambda2}) have three remarkable properties. 

i) At the band center $p_0=0$ the standing waves become unstable for $\beta_0>0$, i.e. for
\be
b_2(h_rh_i-c_rc_i)+h_r^2-c_r^2>0, \label{e:center}
\ee
as already found in \cite{CoFa85}. It was, however, not noted in that early work
that this implies that near the transition from 
standing waves to traveling waves, $h_r \rightarrow c_r^-$, the solution at the band center
becomes unstable for
\be
b_2c_i< b_2h_i,  \label{e:reduced}
\ee
i.e. the condition involves only the imaginary parts $b_2$, $c_i$,
and $h_i$, and even very weak dispersion can be
sufficient to destabilize the standing waves in this regime. 
This is to be contrasted with the result for traveling waves for which the solution 
in the band center becomes unstable only if the dispersion is sufficiently strong, 
as expressed by the Newell criterion $b_2c_i < -1$ \cite{Ne74,StPr78}.

ii) The second striking feature is that the
 parabola giving the stability limit, $\mu=\mu^{(eck)}(p_0)$,
is shifted with respect to the neutral curve as sketched in fig.\ref{f:oeps}. This is
due to the effect of the group velocity $u$. As a consequence, the instability
of the solution at the band center does not imply the instability of all wave numbers (as 
one might have surmised in analogy with the case of traveling waves). Instead, 
when the solution with $p_0=0$ becomes unstable (cf. (\ref{e:center})), 
the apex of the parabola diverges to $\mu\rightarrow \pm \infty$ and the parabola merely
flips over, becoming convex from above. Thus, an off-center band of wave numbers remains that 
is stable with respect to this instability (cf. fig.\ref{f:oeps}b). Alternatively,
when 
\be
K^{(eck)}=0
\ee
the parabola degenerates to a straight line and flips over in that way. 
In either case, 
for $u \ne 0$,
there is always a range of wavenumbers 
(in particular for either small positive or small negative $p_0$)
over which the solution is stable with respect
to the Eckhaus instability even on the neutral curve. In certain parameter regimes
those solutions with vanishing amplitude are also stable with respect to the conservative
instability at $O(\epsilon)$ and therefore stable with respect to all long-wave
perturbations. By continuity, however, the solution cannot be completely stable
since the trivial state $F=0$ is unstable. Therefore the long-wave analysis predicts that
an instability has to arise which occurs first at a finite perturbation wave number. This
is confirmed by the numerical evaluation of the dispersion relation $\lambda(p)$ 
(see below).

iii) In contrast to the case of traveling waves and that of steady patterns
near threshold the stable standing-wave solutions do not have to lie inside the parabolic
Eckhaus stability limit; the case in which the stable solutions are 
{\it outside} the Eckhaus parabola occurs as well. Since $\lambda_2$ can change sign only 
on the Eckhaus parabola and the neutral curve, the condition that the waves are unstable 
outside the parabola is given by $\beta_2 <0$, i.e.
\be
2b_2h_r(c_ih_r-c_rh_i)-c_r(c_r^2-h_r^2+c_i^2-h_i^2)<0.\label{e:pquadinf}
\ee
 In the absence of dispersion this is always the
case (since $c_r>h_r$). For sufficiently strong dispersion, however, $\lambda_2$ can
become negative for large $p_0$, and the waves are {\it stable outside} the parabola. Comparison with (\ref{e:plininf}) shows that near the
transition from standing waves to traveling waves, i.e. for $h_r \rightarrow c_r^{-}$, 
the behavior of the two instabilities is complementary at large $|p_0|$: either the
pattern is unstable due to $\lambda_1$ or due to $\lambda_2$. However, away from this
transition {\it both} instabilities can arise for large $p_0$ or {\it neither}. 
The latter case is
particularly interesting since it implies that the pattern is stable with respect to 
long-wave perturbations even on the neutral curve. This parameter regime is limited by the 
conditions $\gamma_1=0$ and $\beta_2=0$. Solving for $b_2$ and $h_i$ one obtains
\be
b_2^{(crit)}=-\frac{c_r}{2h_r}\frac{h_i^2-c_i^2+h_r^2-c_r^2}{h_rc_i-c_rh_i}\label{e:b2crit}
\ee
and
\be
h_i^{(1,2)}=(1/c_r)\left(c_ih_r\pm(c_r-h_r)|c|\right) \qquad \mbox{ or } \qquad
h_i^{(3,4)}=(1/c_r)\left(c_ih_r\pm(c_r+h_r)|c|\right). \label{e:hicrit}
\ee
This leads to three intervals in $h_i$ over
 which $\lambda_1^2<0$ 
and $\beta_2=0$. For $h_r>0$ they are given by $(-\infty,h_i^{(4)})$, 
$(h_i^{(2)},h_i^{(1)})$ and $(h_i^{(3)},\infty)$. For $h_r<0$ the values  
$h_i^{(1,2)}$ and $h_i^{(3,4)}$ interchange their roles. In these intervals and
for $b_2$ close to $b_2^{(crit)}$ the long-wave stability analysis suggests that the
pattern is stable even on the neutral curve. Again, this indicates the appearance of
an instability at finite wavelength (see below).  
 
Thus, a number of qualitatively different situations can arise. For
\be
K^{(eck)} > 0
\ee
the parabola is convex from below and otherwise convex from above. In either case the 
solution in the band center is stable for
\be
\beta_0 < 0.
\ee
If the parabola is convex from below this implies that the solution is stable inside
the parabola, otherwise it is stable outside the parabola. In the former case it
is worthwhile to consider the width of the Eckhaus-parabola. For 
\be
\beta_0 \beta_2 >0
\ee
it is narrower than the neutral curve and wider otherwise. This implies that
$\beta_2$ determines the Eckhaus-stability of the solution with large $|p_0|$ (cf. 
(\ref{e:pquadinf})). In each of the cases the conservative 
instability at $O(\epsilon )$ 
can arise inside or outside the interval $(p^{(1)}_{0,1},p^{(1)}_{0,2})$ as determined
by $\gamma_1$. 

\begin{figure}[htb]
\begin{picture}(120,180)(0,0)
\put(-55,-45) {\includegraphics{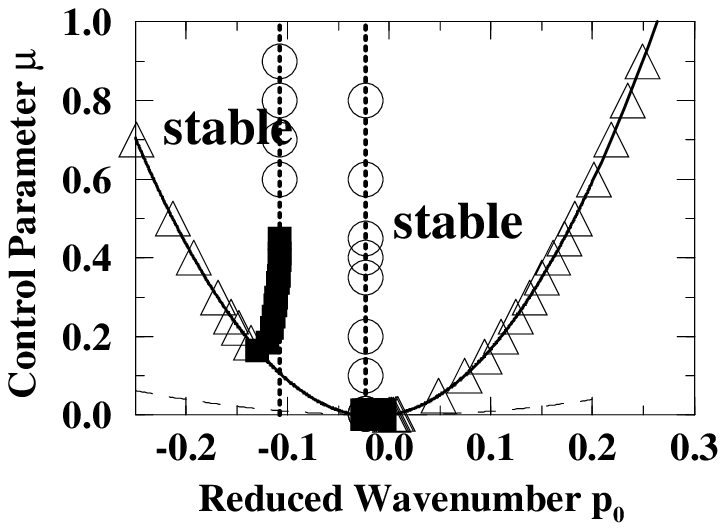}}
\put(190,-45) {\includegraphics{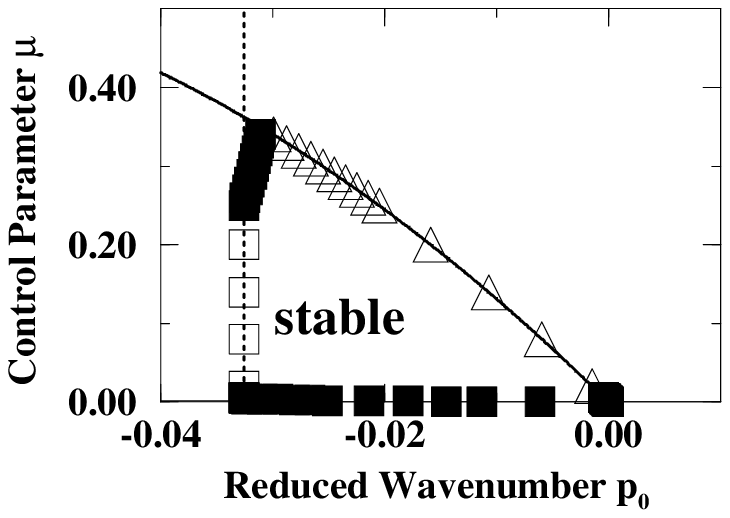}}
\end{picture}
\caption{Stability limits for standing waves. Long-wave expansion at \protect{$O(\epsilon)$}
(dotted line)
and \protect{$O(\epsilon^2)$} (solid line). The dashed line denotes the neutral
curve. The symbols denote the stability limits as obtained from the 
numerical analysis of the dispersion relation 
\protect{$\lambda (p)$} with the solid squares marking
an instability at finite perturbation wave number. 
The parameters are as follows: 
a) $c=1-0.4i$, $h=0.6+0.3i$, $b_2=0.4$, $u=0.1$, 
b) $c=1-0.6i$, $h=0.7+0.6i$, $b_2=0.6$, $u=0.3$.
\protect{\label{f:stabi1}}
}
\end{figure}

\begin{figure}[htb]
\begin{picture}(120,180)(0,0)
\put(-55,-45) {\includegraphics{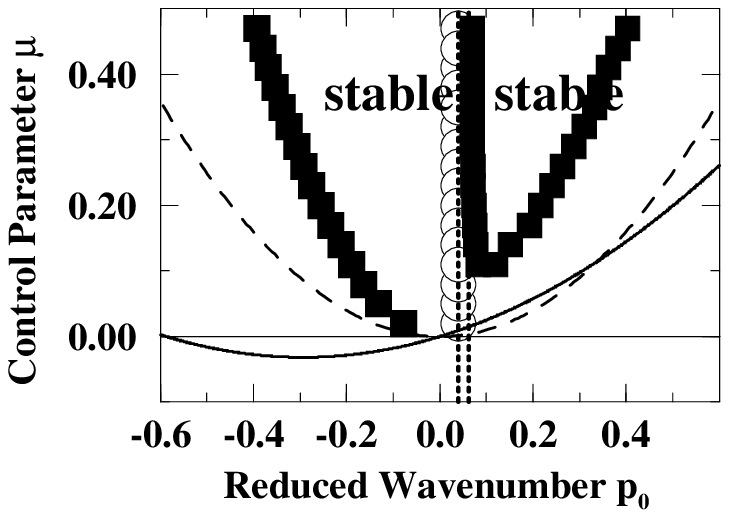}}
\put(190,-45) {\includegraphics{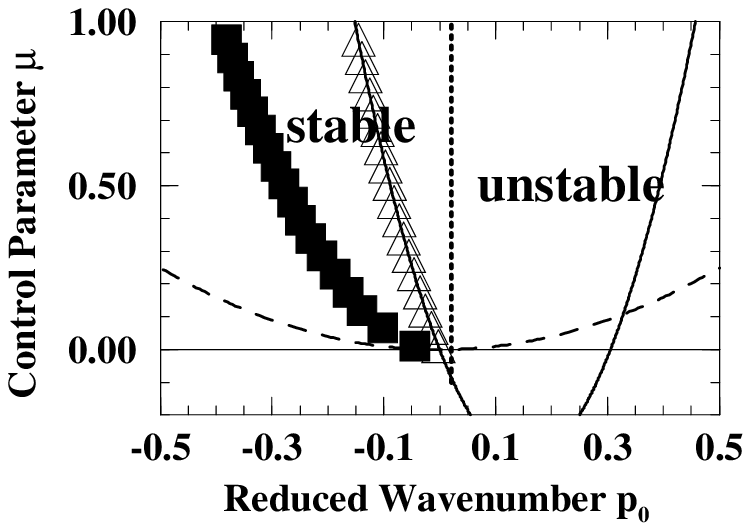}}
\end{picture}
\caption{Stability limits for standing waves (cf. fig.\protect{\ref{f:stabi1}}).
The parameters are as follows: 
a) $c=1+0.3i$, $h=0.5+0.5i$, $b_2=-1.96$, $u=0.2$,
b) $c=1+0.3i$, $h=0.2+0.5i$, $b_2=-4.85$, $u=0.2$,
\protect{\label{f:stabi2}}
}
\end{figure}

\begin{figure}[htb]
\begin{picture}(120,180)(0,0)
\put(-55,-45) {\includegraphics{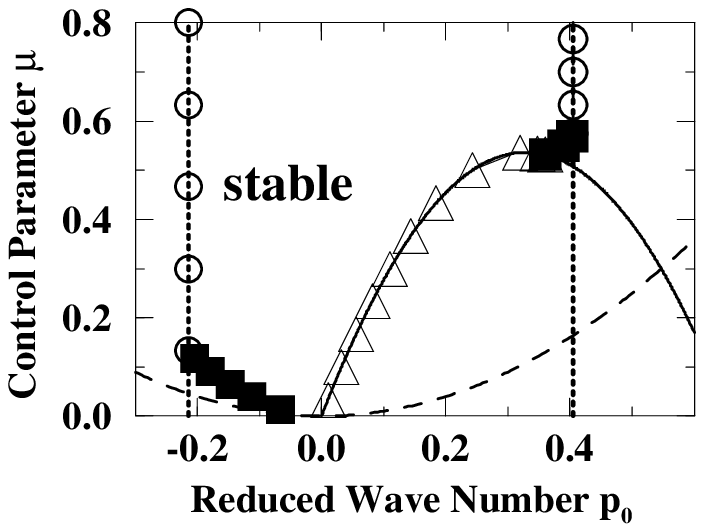}}
\put(190,-45) {\includegraphics{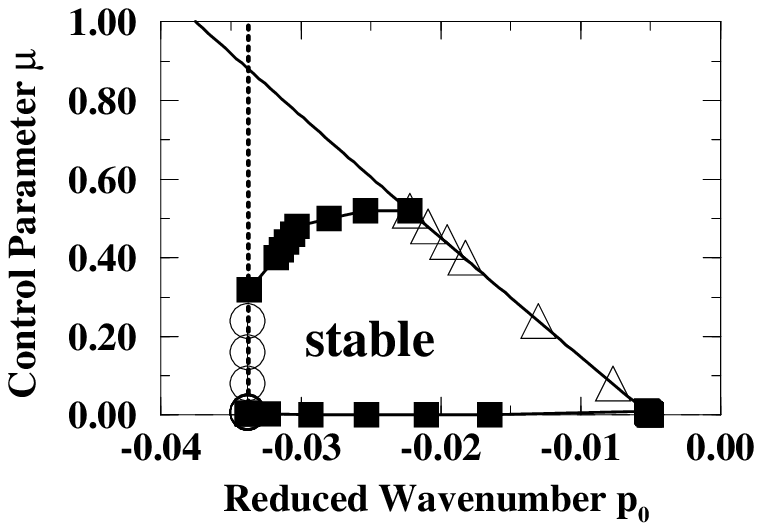}}
\put(340,23) {\includegraphics{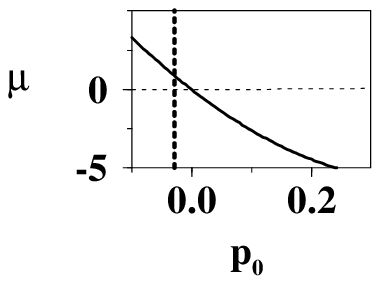}}
\end{picture}
\caption{Stability limits for standing waves (cf. fig.\protect{\ref{f:stabi1}}). 
The parameters are as follows: 
a) $c=1-0.3i$, $h=0.7+0.7i$, $b_2=-1$, $u=1$,
b) $c=1+0.7i$, $h=0.9+1.5i$, $b_2=0.7$, $u=0.5$.
\protect{\label{f:stabi3}}
}
\end{figure}

In figs.\ref{f:stabi1}-\ref{f:stabi3} representative cases for the stability limits are
shown as obtained from the long-wave results (\ref{e:oeps}) 
and (\ref{e:lambda2}), and
 by numerically solving the dispersion relation $\lambda(p)$. 
In fig.\ref{f:stabi1}a
the parabola is convex from below
 and the Eckhaus-stable solutions are inside. For the parameters chosen,
the conservative instability destabilizes the solutions in a vertical strip inside
that parabola. When the parameters are changed so as to render the band center unstable
the parabola  becomes convex from above and 
one obtains the situation given in fig.\ref{f:stabi1}b (Note the enlarged
wavenumber scale). Only the small area between the 
parabola and the vertical stability limit corresponds to stable solutions. 
Since the long-wave instabilities do not arise on the neutral curve for small negative
$p_0$ an instability at finite wavelength arises there. It is indicated by solid squares.

In fig.\ref{f:stabi2}a the Eckhaus parabola is wider than the neutral
curve. Since the $O(\epsilon)$-instability destabilizes the solution only
in the small range close to the band center the wavenumber band is
limited by the finite-wavelength instability for large $|p_0|$. 
In fig.\ref{f:stabi2}b the stable solutions are outside the Eckhaus parabola
and again perturbations at finite wavelength are important. 

For the 
parameters chosen in fig.\ref{f:stabi3}a the stable solutions are also
outside the parabola, but
the parabola is now convex from above. 
In the final case, shown in fig.\ref{f:stabi3}b, the standing waves are again 
Eckhaus-stable outside the parabola, which is convex from below 
(cf. inset). Now
the conservative instability limits the wavenumber band on the other side 
(compare with fig.\ref{f:stabi2}b) and only a very small region of stability remains. 

As indicated above, due to the finite group velocity there is 
generically a range of wave numbers
for which the solution is stable on the neutral curve with respect to 
long-wave perturbations. Figs.\ref{f:stabi1}-\ref{f:stabi3} confirm that therefore
generically instabilities occur
at finite perturbation wave number $p$. This is to be contrasted with the case of
traveling waves where they appear only for sufficiently strong dispersion 
\cite{StPr78,MaVo93}. 

\begin{figure}[htb]
\begin{picture}(120,350)(0,0)
\put(-55,130) {\includegraphics{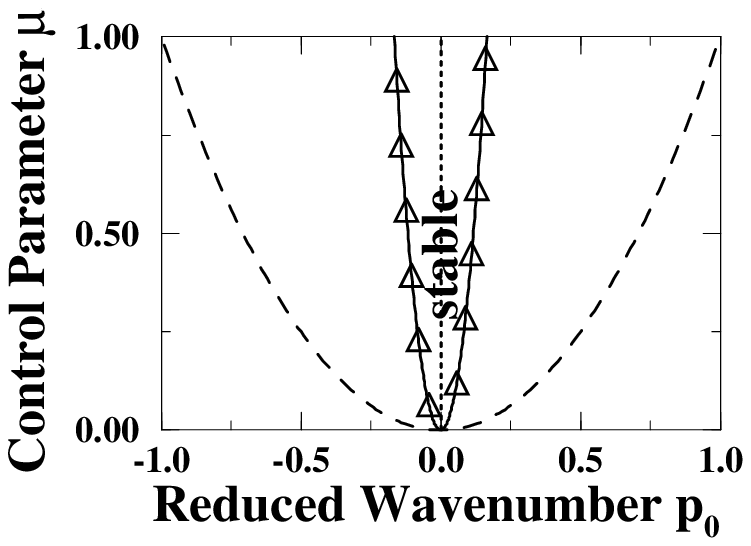}}
\put(160,130) {\includegraphics{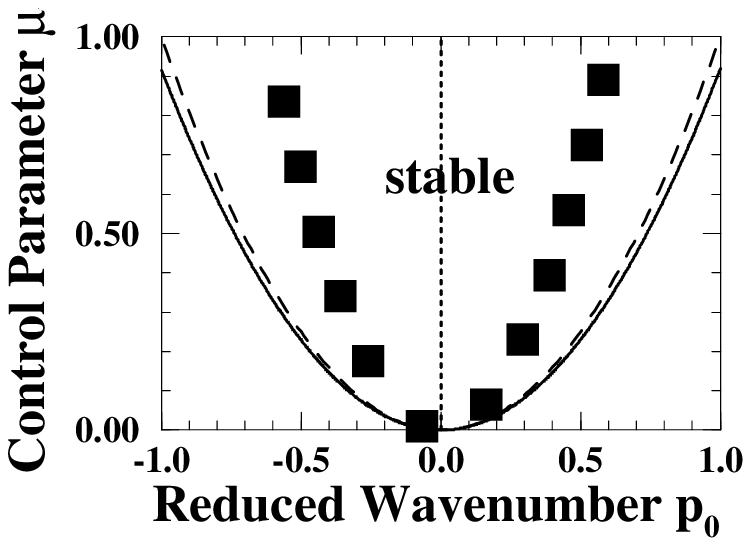}}
\put(-55,-40) {\includegraphics{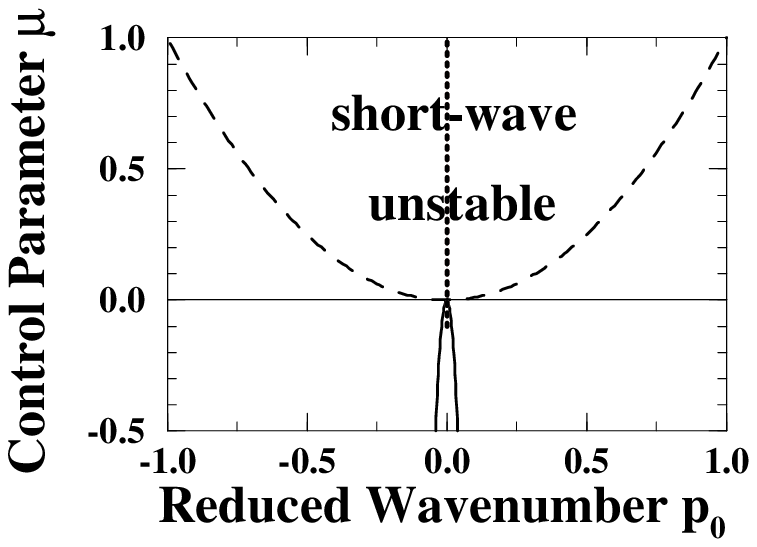}}
\end{picture}
\caption{Stability limits for standing waves with vanishing group velocity. 
Long-wave expansion at \protect{$O(\epsilon)$}
(dotted line)
and \protect{$O(\epsilon^2)$} (solid line), 
and instability at finite perturbation wave number (solid squares). 
The symbols denote the stability limits as obtained from the 
numerical analysis of the dispersion relation 
\protect{$\lambda (p)$}. Neutral curve given by dashed line.
Note that in c) {\it all} wavenumbers are {\it stable} according to the long-wave analysis, 
but they are in fact all {\it unstable} due to an instability at finite $p$.
The parameters are as follows:
a) $c=1+1.1i$, $h=0.6+0.5i$, $b_2=-0.5$, $u=0$,
b) $c=1+1.1i$, $h=0.6+0.5i$, $b_2=9$, $u=0$,
\protect{\label{f:u0}}
}
\end{figure}

Let us briefly discuss the case of vanishing group velocity separately. 
As is apparent from (\ref{e:oeps}) and (\ref{e:plin})
the stability with respect to long-wave perturbations does not depend on the 
wavenumber $p_0$ to $O(\epsilon)$: either all wavenumbers are 
unstable ($\gamma_1<0$) or stable ($\gamma_1>0$).
In the latter case the long-wave stability limits are determined by (\ref{e:eckhaus}).
At $O(\epsilon^2)$ one obtains a symmetric Eckhaus parabola. Again it can be convex
down or up and can be wider or narrower than the neutral curve. Two cases are shown 
in fig.\ref{f:u0}a,b. In fig.\ref{f:u0}a the Eckhaus parabola is narrower than the 
neutral curve and the stability limit is indeed given by the long-wave instability.
In fig.\ref{f:u0}b the parabola is wider and the stability is limited by a 
instability with finite $p$. 

For $u=0$ the instability at finite $p$ can be captured analytically
near the band center because it arises there for $p=O(p_0)$.
To that end 
the perturbation wavenumber $p$ and the wavenumber $p_0$ are taken to be
of the same order, i.e.
\be
p=\epsilon p_1, \qquad p_0=\epsilon p_{01}, \qquad 
\lambda = \epsilon \lambda_1 + \epsilon^2 \lambda_2 + ...
\ee
At $O(\epsilon^2)$ one obtains $\lambda_1=0$. Note that (\ref{e:lambda2}) does therefore
not apply and one has to go to higher order. At $O(\epsilon^4)$ one finds
\be
\lambda_2^{(1,2)} = p_1^2 \beta_0 \pm p_1 \sqrt{W_1}
\label{e:lambda2P0}
\ee
with
\be
W_1=\frac{b_2^2(c_rh_i-c_ih_r)^2}{(c_r^2-h_r^2)^2}p_1^2-4\left(b_2^2-2\frac{b_2(c_rc_i-h_rh_i)}
{c_r^2-h_r^2}+
\frac{c_i^2-h_i^2}{c_r^2-h_r^2}\right)p_{01}^2.\label{e:W_1}
\ee
The limit 
$p_0 \to 0$ of the previous result (\ref{e:oeps}),(\ref{e:lambda2}) for $\lambda_{1,2}$ is regained  by taking the limit $p_1\rightarrow 0$ 
in  (\ref{e:lambda2P0},\ref{e:W_1}).
Thus, if the first term in (\ref{e:lambda2P0}) is negative 
and the prefactor of $p_{01}^2$ in
(\ref{e:W_1}) is positive $\sqrt{W_1}$ is imaginary  in the long-wave limit and
 the waves with small $p_0$ are stable.
However, the first term in $W_1$ becomes important when the modulation wavenumber 
becomes larger. Thus, for $|p_1| \gg |p_{01}|>0$ (\ref{e:lambda2P0}) yields an instability
at a finite modulation wavenumber if $\beta_0 + \sqrt{W_1(p_{01}=0)} > 0$, i.e.
\be
b_2(c_i-h_i)+c_r-h_r<0, \qquad \mbox{ or } \qquad b_2(c_i+h_i)+c_r+h_r<0.\label{e:stabu0p0}
\ee
Such a case is shown in fig.\ref{f:u0}c in which all wavenumbers are stable
with respect to long-wave perturbations ($\beta_0<0$, $\beta_2>0$, and
$\gamma_1>0$), but all wavenumbers are {\it unstable} to perturbations at
finite $p_1$. Note, that the relevant perturbation wavenumber goes to 0 in the band center.
In the special case $p_0=0$ the stability conditions  (\ref{e:stabu0p0}) are
therefore also obtained in a long-wave analysis \cite{Sa95}.


\section{Analysis for Large Group Velocity}
\label{s:knob}
Recently there has been some discussion of the fact that equations 
(\ref{e:swA},\ref{e:swB}) are not 
obtained in that form when a direct expansion of the microscopic equations like the Navier-Stokes
equations is performed near threshold \cite{Kn95,KnDe90,Kn92,MaVo92,Ve93,MaVe96,MaVe98}. 
Since the group velocity of the waves is
generically of O(1) one is, strictly speaking, either led to consider a hyperbolic 
system without the second space derivatives in (\ref{e:swA},\ref{e:swB}) \cite{MaVe96,MaVe98}
or to
introduce two different time scales $T_1=\epsilon t$
 and $T_2=\epsilon^2t$ \cite{Kn95,KnDe90,Kn92,MaVo92,Ve93}. 
One then obtains already at quadratic order a solvability condition,
\bea
\partial_{T_1}A+u\partial_Y A = 0,\label{e:nlA}\\
\partial_{T_1}B-u\partial_Y B = 0\label{e:nlB}.
\eea
Solving these equations in the frames co-moving with the respective waves ($Y_\pm=Y\mp uT_1$)
one then obtains at third order
\bea  
\partial_{T_2} A &=& \mu A + ( 1+ ib_2) \partial_{Y_+}^2 A 
 -c|A|^2A - h<|B|^2>A,\label{e:swAnl} \\
\partial_{T_2} B &=& \mu B + ( 1+ ib_2) \partial_{Y_-}^2 B
 -c|B|^2B - h<|A|^2>B,\label{e:swBnl}
\eea
with $<...>$ denoting spatial averages. 
These equations differ from  (\ref{e:swA},\ref{e:swB}) in particular
in their non-local cross-coupling terms between
$A$ and $B$:  At the band center 
the linear stability with respect to long-wave perturbations
 of the standing waves has been investigated within these equations and
a result quite different from that obtained in sec.\ref{s:sw} has been obtained 
\cite{Kn92}. It has been
pointed out that the difference between the two long-wave results is due to the fact that
two limits have been taken: small amplitudes, i.e. $\mu \rightarrow 0$, and long waves,
i.e. $p \rightarrow 0$. These two limits do not commute. It has been argued that the 
asymptotically correct sequence is to take first $\mu \rightarrow 0$, which leads to the
non-local equations (\ref{e:swAnl},\ref{e:swBnl}), 
and then $p \rightarrow 0$. It is therefore
of great interest to understand the connection between the two results and in particular
the respective regimes of relevance. 
As long as no long-wave limit is taken the local (reconstituted) equations are expected
to be valid.

As discussed in \cite{Kn92} the stability result obtained from the non-local equations 
can be recovered from the local equations if the limit $\mu \rightarrow 0$ is considered before
taking the long-wave limit. To extend the analysis presented in \cite{Kn92} away from the
band center we consider the full $4^{th}$-order dispersion relation of perturbations of 
standing waves within the local equations taking the limit $\mu \rightarrow 0$ before 
considering the long-wave limit.
Due to the parabolic character of the neutral curve the wave numbers have to be 
rescaled along with $\mu$,
\be
\mu = \eta^2 \mu_2, \qquad p_0 = \eta p_{01}, \qquad p = \eta p_1, \qquad F = \eta F_1. \label{e:etaexp}
\ee
The growth rates are also rescaled,
\be
\lambda = iup +\eta^2 \Lambda.
\ee
One then obtains at $O(\eta^6)$ 
\bea
\Lambda^2 + \Lambda \left( 2p_1^2-4ip_{01} b_2 p_{1} + 2 c_r F_1^2 \right) 
-(1+b_2^2)(4 p_{01}^2 - p_{1}^2) p_{1}^2 \nonumber \\
-4i(c_r b_2-c_i)p_{1} p_{01} F_1^2  +2 (c_r+c_i b_2) p_{1}^2 F_1^2  =0
\eea
Note that $u$, which is assumed to be $O(1)$, does not affect the (convective) destabilisation.
Actually this dispersion relation coincides with that obtained for traveling waves, if $F_1$
is replaced by the traveling-wave amplitude.
If one now expands in small perturbation wavenumbers $p_1$, one obtains for the
growth rate
\be
\Lambda = 2ip_{01} p_{1} \frac{-c_i+c_r b_2}{c_r} + 
p_{1}^2 \left( 2 \frac{|c|^2}{c_r^3 F_1^2}p_{01}^2 - 
\frac{c_r+c_i b_2}{c_r} \right) +O(p_{1}^3). \label{e:Lanl}
\ee
Thus, in this limit no conservative instability arises and 
the stability boundaries are given by a parabolic curve.
Note that the term in brackets corresponds to the negative of the expression in Eq.(\ref{e:travstab}).
In fact, the destabilization of all periodic solutions is indicated by the
loss of stability of the solution in the band center which occurs for the same parameter
values (as for the traveling wave),
\be
b_2 c_i < -c_r.\label{e:newell}
\ee
This is in strong contrast to the result found in (\ref{e:lambda2}) 
above, where for non-vanishing group velocity
the Eckhaus curve was shifted with respect to the neutral curve. 
None of the complex behavior 
discussed above is obtained in (\ref{e:swAnl},\ref{e:swBnl}).

Central to the  difference between the two 
stability results is the long-wave limit. To understand the
connection between them it is useful to consider a distinguished limit in which
$\mu$ and the perturbation wave number $p$ are both small of the same order,
\bea
\mu = \epsilon^2 \mu_2, \qquad p_0 = \epsilon^{2} p_{02}, \qquad p = \epsilon^2 p_2,
\qquad F = \epsilon F_1, \qquad \lambda = iup + \epsilon^4 \lambda_4, \label{e:epsilonexp}
\eea
which differs from the previous limit (\ref{e:etaexp})
in the scaling of the modulation wavenumber $p$ and that
of the growth rate $\lambda$. It turns out that the wavenumber $p_{0}$ of 
the standing waves themselves has to be taken to be of 
$O(\epsilon^{2})$ in order to obtain a single expression that reduces 
to the results of the local and of the non-local equations in 
suitable limits.
 At leading order ($O(\epsilon^9)$) one obtains then
\bea
\lambda_{4}&=&{\cal N}^{-1}\, 
p_{2}^{2}\left\{2F_{1}^{2}h_{r}(c_{r}h_{i}-c_{i}h_{r})\,up_{02}-
(c_{r}^{2}-h_{r}^{2})
\left(c_{r}^{2}-h_{r}^{2}+b_{2}(c_{r}c_{i}-h_{r}h_{i})\right)F_{1}^{4}
-u^{2}c_{r}(c_{r}+b_{2}c_{i})p_{2}^{2}\right\}+\nonumber \\
&+&i{\cal N}^{-1}p_{2}\left\{2(c_{r}^{2}-h_{r}^{2})
\left(b_{2}(c_{r}^{2}-h_{r}^{2})+h_{r}h_{i}-c_{r}c_{i}\right)F_{1}^{4}\,p_{02}+
up_{2}^{2}\left(2up_{02}c_{r}
(c_{r}b_{2}-c_{i})-b_{2}h_{r}(c_{r}h_{i}-c_{i}h_{r})F_{1}^{2}\right)\right\}
\label{e:lambda4}\\
{\cal N} &=& (c_{r}^{2}-h_{r}^{2})F_{1}^{4}+u^{2}p_{2}^{2}c_{r}^{2}.
\eea
For small amplitudes $F_{1}$ (\ref{e:lambda4}) reduces to
\be
\lambda_{4}=-p_{2}^{2}\frac{c_{r}+b_{2}c_{i}}{c_{r}}+
2p_{02}h_{r}\frac{c_{r}h_{i}-c_{i}h_{r}}{uc_{r}^{2}}F_{1}^{2}+\nonumber \\
ip_{2}\left\{2p_{02}\frac{c_{r}b_{2}-c_{i}}{c_{r}}-
\frac{h_{r}b_{2}(c_{r}h_{i}-c_{i}h_{r})}{uc_{r}^{2}}F_{1}^{2}\right\}+O(F_{1}^{4}).
\ee
This limit agrees to leading order with the result from the non-local 
theory (\ref{e:Lanl}). Note that due to the choice 
$p_{0}=O(\epsilon^{2})$ the wavenumber dependence of $\lambda$ in the 
non-local theory (cf. (\ref{e:Lanl})) does 
not appear to this order. In the limit of large amplitudes one 
obtains from (\ref{e:lambda4}) 
\bea
\lambda_{4}&=&p_{2}^{2}\left\{\beta_{0}-\beta_{1}\frac{p_{02}u}{F_{1}^{2}}\right\} + 
\nonumber \\
&&ip_{2}\left\{2p_{02}\frac{b_{2}(c_{r}^{2}-h_{r}^{2})+h_{i}h_{r}-c_{i}c_{r}}
{c_{r}^{2}-h_{r}^{2}}+up_{2}^{2}
\frac{h_{r}b_{2}(c_{r}h_{i}-c_{i}h_{r})}{(c_{r}^{2}-h_{r}^{2})^{2}F_{1}^{2}}\right\}
+O(F_{1}^{-4})
\eea
with $\beta_{0}$ and $\beta_{1}$ defined in (\ref{e:alpha13}). Comparison 
with (\ref{e:oeps}) and (\ref{e:lambda2}) shows that the limit 
(\ref{e:epsilonexp}) also 
reproduces to leading order the result from the local theory. 
The imaginary part corresponds to an expansion of $\lambda_{1}$ from 
 (\ref{e:oeps}) for small $p_{0}$, noting that a term $iup_{2}$ has been 
 split off in (\ref{e:epsilonexp}).
 
\begin{figure}[htb]
\begin{picture}(120,350)(0,0)
\put(-55,130) {\includegraphics{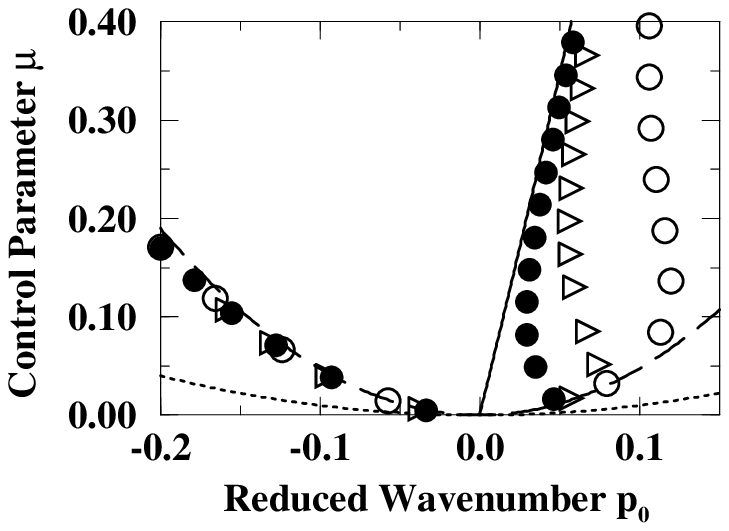}}
\put(160,130) {\includegraphics{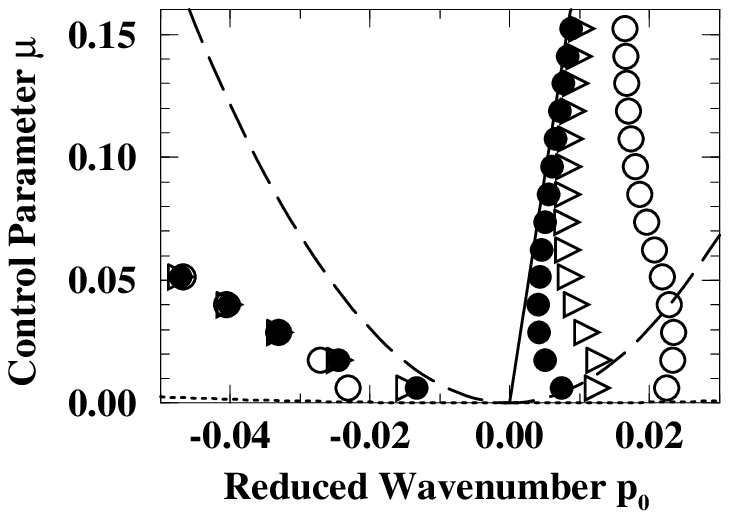}}
\put(-55,-40) {\includegraphics{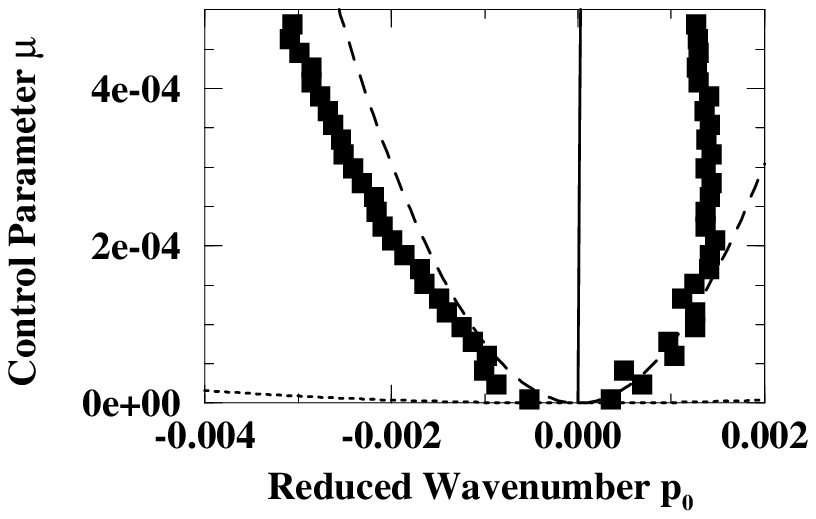}}
\end{picture}
\caption{Comparison between stability results from non-local 
(\protect{\ref{e:swAnl}},\protect{\ref{e:swBnl}}) and local equations
(\protect{\ref{e:swA}},\protect{\ref{e:swB}}). 
Solid line from (\protect{\ref{e:eckhaus}}), dashed line from (\protect{\ref{e:Lanl}}),
dotted line gives the neutral curve. 
Symbols denote the stability limits from the full dispersion 
relation for  $p_{min}=0.02$ (open circles), $p_{min}=0.01$ (triangles),
$p_{min}=0.005$ (solid circles), and $p_{min}=0.0005$ (in c)).
The parameters are $c=1+0.5i$, $h=0.5+0.5i$, $u=10$, and $b_2=0$ (in a))
and $b_2=-1.9$ (in b) and c)).
\protect{\label{f:loc-nl}
}
}
\end{figure}

 Thus, the distinguished limit (\ref{e:epsilonexp}) provides a 
 connection between the two regimes and allows to give an expresssion 
 for the parameters at which the  cross-over occurs. Focussing on the 
 band center, $p_{0}=0$, the stability is determined by the second and 
 third term in (\ref{e:lambda4}). A reasonable definition for the 
 cross-over is therefore their ratio $\Gamma$,
 \be
 \Gamma = 
\left| \frac{(c_{r}^{2}-h_{r}^{2})(c_{r}^{2}-h_{r}^{2}+b_{2}(c_{r}c_{i}-h_{r}h_{i}))}
 {c_{r}(c_{r}+b_{2}c_{i})}\right| \,\frac{F^{4}}{u^{2}p^{2}}.
 \label{e:Gamma}
 \ee
The non-local theory is appropriate for $\Gamma \ll 1$.  
For fixed values of the group velocity $u$, this corresponds to small amplitudes  
$F^{2}\sim \mu$.
The cross-over to the regime in which the results of the local theory
are recovered occurs near $\mu_c$, which is defined $via$ $\Gamma=1$.
Eq.(\ref{e:Gamma}) shows that $\mu_c$ depends not only on the group velocity $u$ 
but also on the wavenumber $p$ of the relevant perturbations and
therefore on the size $L=2\pi/p_{min}$  of the physical system under consideration,
\be
\mu_c \propto u p_{min}.\label{e:cross-over}
\ee
This dependence is illustrated in fig.\ref{f:loc-nl}a. There the 
analytical results (\ref{e:eckhaus},\ref{e:Lanl}) are shown as well as the
numerical result from the full dispersion relation
for three system sizes corresponding to $p_{min}=0.02$, $p_{min}=0.01$, and $p_{min}=0.005$. 
The other parameters
are $c=1+0.5i$, $h=0.5+0.5i$, $u=10$, and $b_2=0$. As expected, the non-local theory 
is correct for small $\mu$. For larger $\mu$ strong deviations arise for $p_0 >0$.
They signify the cross-over to the result of the local equations. In agreement with
(\ref{e:cross-over}), in longer systems the cross-over occurs at smaller values of $\mu$.
This illustrates the fact that the difference between the two approaches is due to the
fact that the limits $\mu \rightarrow 0$ and $p \rightarrow 0$ are not interchangeable.
In the limit $p \rightarrow 0$ at fixed $\mu$ one obtains the result of the local equations,
whereas for $\mu \rightarrow 0$ at fixed $p$ the result of the
non-local equations are obtained. It is noteworthy, that the cross-over occurs at quite
small values of $\mu$ even though the group velocity is quite large ($u=10$). 
At the same time the instability for $p_0 <0$ is not captured at all 
within  the long-wave limit of the local equations. As discussed in 
sec.\ref{s:sw}, within the long-wave
limit the waves would be stable all the way to the neutral curve, implying that
a short-wave instability arises. Interestingly, although this instability appears in 
the local equations as a short-wave instability, it is captured correctly 
in the long-wave analysis of the non-local equations.
 

Eq.(\ref{e:Gamma}) suggests that the cross-over to the result of the local equations
occurs at a lower value of $\mu$ if $c_r+c_ib_2$ is small, i.e. close to the Newell  
criterion (\ref{e:newell}) for the standing waves (within the non-local
equations). This is demonstrated in 
fig.\ref{f:loc-nl}b. There the stability limits are shown for the same parameters as
in fig.\ref{f:loc-nl}a except for $b_2$ which is here $-1.9$. Thus, $c_r+c_i b_2 = 0.05$. 
As in the case $b_2=0$, the long-wave results of the local theory 
are adequate for long systems and larger $\mu$ (note the
different scales as compared to fig.\ref{f:loc-nl}a) for $p_0>0$. 
The long-wave
stability limits obtained from the non-local equations deviate, however, quite strongly
even at very small values of $\mu$. 
Strikingly, before the parabolic dependence of the non-local result
takes over when $\mu$ is decreased, additional deviations arise, which 
are due to the terms of $O(p^4)$ in (\ref{e:Lanl}). They are
important
for $\mu \leq \mu_f \propto p^2$. Of course, 
if the system size is increased these deviations become smaller, but at the
same time the cross-over is shifted to lower values
$\mu_c$, as well. More precisely, one has the scaling 
\be
\mu_f \propto p^2 \ll \mu_c \propto p \qquad \mbox{for } p\ll 1.
\ee
Thus, for sufficiently large systems there always remains
an intermediate regime in which the parabolic behavior from the
non-local equations is relevant. It can, however, occur for values of $\mu$ that 
are extremely small. This is demonstrated in fig.\ref{f:loc-nl}c which shows a 
blow-up\footnote{For these parameter values round-off errors produced some scatter in the data.} 
of fig.\ref{f:loc-nl}b for $p_{min}=0.0005$. The usual local Ginzburg-Landau equations
contain this result and capture in addition the
rich possibilities that can arise for larger values
of $\mu$. 

\begin{figure}[htb]
\begin{picture}(120,180)(0,0)
\put(-55,-40) {\includegraphics{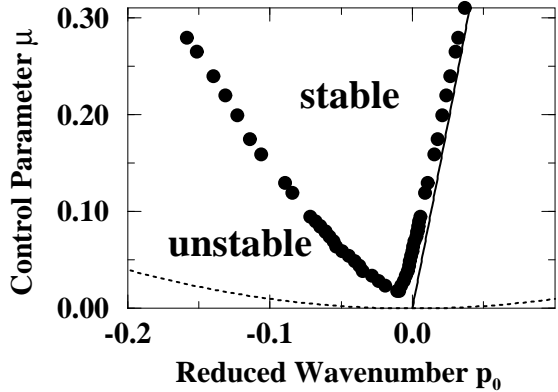}}
\end{picture}
\caption{Comparison between stability results from non-local 
(\protect{\ref{e:swAnl}},\protect{\ref{e:swBnl}}) and local equations
(\protect{\ref{e:swA}},\protect{\ref{e:swB}}). Within the non-local equations all
standing waves are unstable. 
Solid line from (\protect{\ref{e:eckhaus}}), 
dotted line gives the neutral curve. 
Symbols denote the stability limits from the full dispersion 
relation for $p_{min}=0.005$ (solid circles).
The parameters are $c=1+0.5i$, $h=0.5+0.5i$, $u=3$, and $b_2=-2.2$ (cf. fig.\protect{\ref{f:loc-nl}}).
\protect{\label{f:BF}
}
}
\end{figure}

A further interesting case is obtained if the Benjamin-Feir criterion $c_r+b_2c_i$ 
is satisfied within the non-local equations. 
In that case all standing waves are unstable immediately at 
onset. At slightly larger values of the control parameter, however, the waves can become
stable. This is shown in fig.\ref{f:BF}. Although with increasing values of the group velocity
the gap of instability increases, it is seen that the restabilization can occur already for
quite small values of the control parameter even for moderately large group velocity.

Of course, at larger values of $\mu$ higher-order corrections will have to be included. 
Since the expansion leading to $(\ref{e:swA},\ref{e:swB})$ is only asymptotic 
(rather than convergent) firm predictions for {\it finite} values of $\mu$ cannot be made. However, if the cross-over to the local equations
 occurs for small values of $\mu$
it is quite reasonable to expect that the rich behavior
displayed in figs.\ref{f:stabi1}-\ref{f:stabi3} will not be strongly affected
by the higher-order corrections. 
  
\section{Numerical Simulations in the Unstable Regimes}

\begin{figure}[htb]
\begin{picture}(120,220)(0,0)
\put(0,10) {\includegraphics{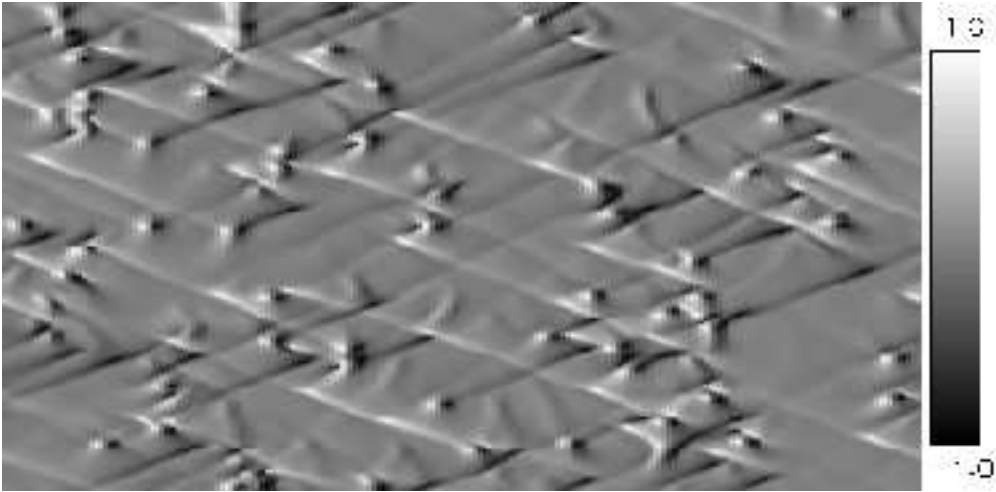}}
\end{picture}
\caption{Space-time diagram (space horizontally, time vertically 
upward) of evolution obtained by
numerical simulation of 
(\protect{\ref{e:swA}},\protect{\ref{e:swB}})
 for $\mu=0.4$, $c=1-0.6i$, $h=0.7+0.6i$, $b_2=0.6$, $u=0.3$ (cf.~fig.\protect{\ref{f:stabi1}b}).
The shades of gray indicate the normalized difference $(|A|-|B|)/(|A|+|B|)$ between the amplitudes
of the right- and left-traveling wave (white corresponds to right-traveling, 
black to left-traveling waves).
The length of the system is $L=400$  and the 
time shown is from $t=4000$ to $t=5000$.
\protect{\label{f:swsim1}}
}
\end{figure}

\begin{figure}[htb]
\begin{picture}(120,220)(0,0)
\put(0,10) {\includegraphics{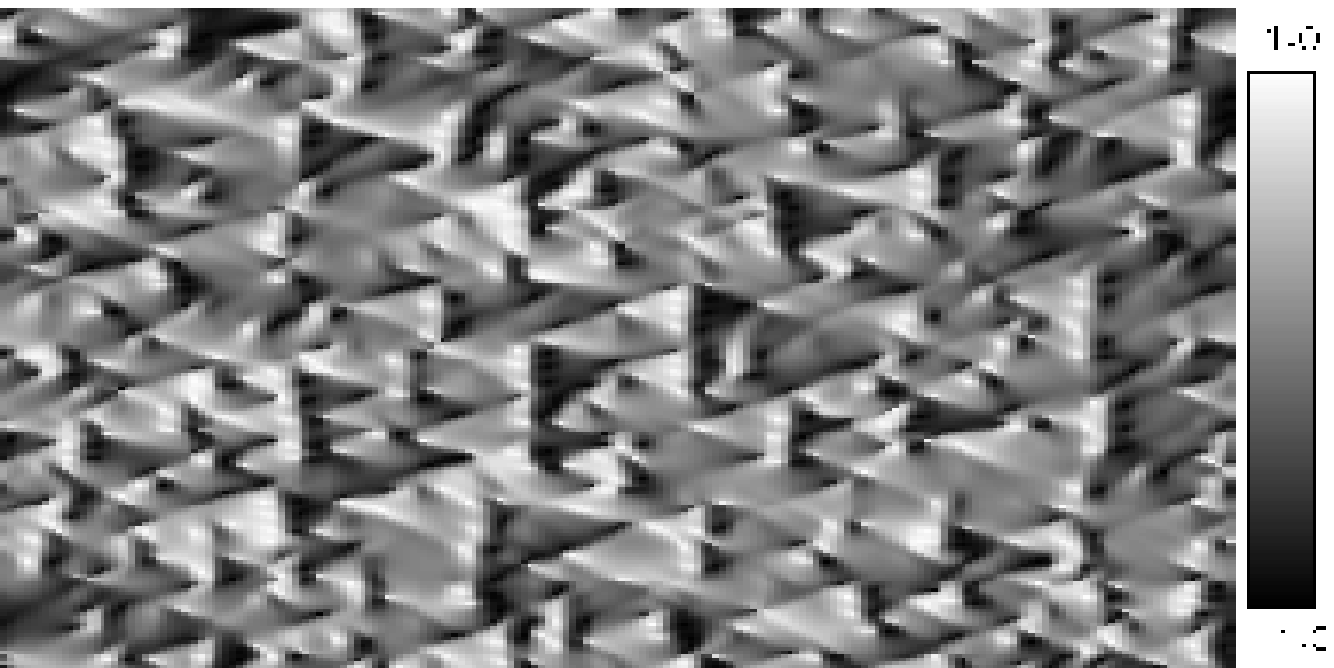}}
\end{picture}
\caption{Space-time diagram (space horizontally, time vertically 
upward) of evolution obtained by
numerical simulation of 
(\protect{\ref{e:swA}},\protect{\ref{e:swB}})  for 
 \protect{$h_r=0.9$}. 
(other parameters as in fig.\protect{\ref{f:swsim1}}).
\protect{\label{f:swsim2}}
}
\end{figure}

\begin{figure}[htb]
\begin{picture}(120,220)(0,0)
\put(0,10) {\includegraphics{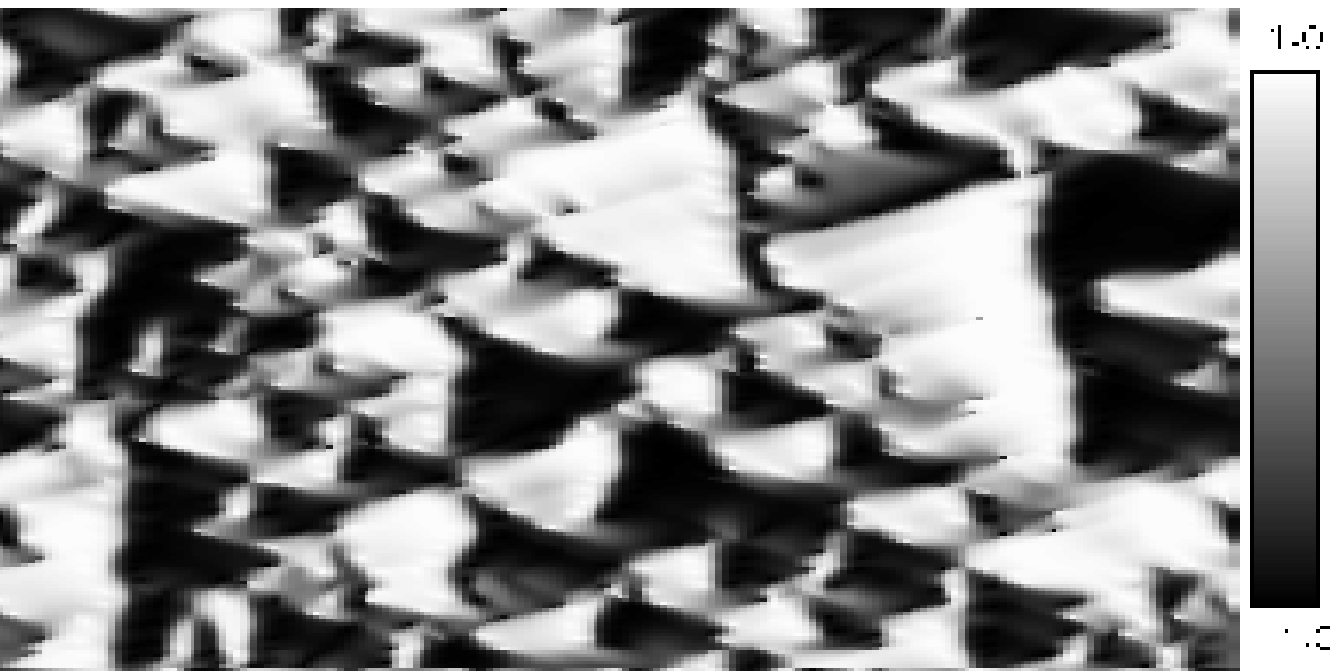}}
\end{picture}
\caption{Space-time diagram (space horizontally, time vertically 
upward) of evolution obtained by
numerical simulation of 
(\protect{\ref{e:swA}},\protect{\ref{e:swB}})  for 
 \protect{$h_r=1.1$}. 
(other parameters as in fig.\protect{\ref{f:swsim1}}).
\protect{\label{f:swsim3}}
}
\end{figure}

In order to determine the dynamics ensuing from the instabilities found analytically we have
solved (\ref{e:swA},\ref{e:swB}) numerically for a range of parameters. We have focussed
in particular on the behavior in the vicinity of the transition from standing waves to
traveling waves ($h_r \rightarrow c_r^-$), since in this regime even weak dispersion can 
lead to a destabilization of standing waves at all wavenumbers (cf. fig.\ref{f:stabi1}b). 
Figures \ref{f:swsim1}-\ref{f:swsim3} 
show a sequence of space-time diagrams for increasing values of $h_r$ straddling the 
transition point $h_r=c_r$.
As initial condition a slightly perturbed standing
wave was chosen. The shades of gray in the space-time 
diagrams indicate the normalized difference $(|A|-|B|)/(|A|+|B|)$ between the amplitudes
of the right- and left-traveling wave with bright areas indicating 
domains in which the right-traveling wave dominates and dark ones 
those in which the left-traveling wave dominates.
 For $h_r=0.7$ both wave components are about equally strong resulting in a
 solution which  corresponds mostly to a standing
wave. This standing wave exhibits, however, complex dynamics due to the persistent
occurrence of phase slips in one of the two wave components and due to fairly
localized  disturbances propagating through the system. 

Closer to the transition
to traveling waves ($h_r=0.9$) the traveling-wave components become stronger 
and a characteristic behavior is seen (fig.\ref{f:swsim2}). 
Domains of left-traveling waves expand rapidly to 
the right for some time until the adjacent right-traveling wave starts to expand rapidly
to the left pushing the left-traveling wave back. In the space-time diagram this
sequence leads to dark triangles pointing to the right and bright ones to the left. 
Even for $h_r=1.1$, i.e. in the regime in which traveling waves of a given wavenumber are stable with respect to standing waves of the same wavenumber, 
these dynamics of rapid expansion and contraction persist as shown in fig.\ref{f:swsim3}. 

\begin{figure}[htb]
\begin{picture}(120,180)(0,0)
\put(0,-40) {\includegraphics{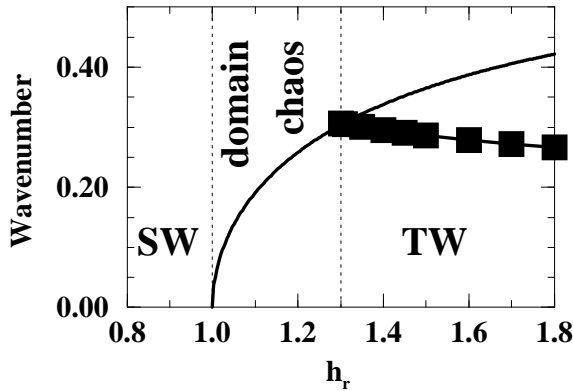}}
\end{picture}
\caption{Wavenumber selected by source (squares) and stability limit of traveling
waves with respect to counter-propagating waves (solid line). Between the dotted lines
domains of traveling waves are unstable. (Parameters as in fig.\protect{\ref{f:swsim1}}).
\protect{\label{f:quell}}
}
\end{figure}

The origin of the complex dynamics shown in fig.\ref{f:swsim3} 
is related to the selection of a non-zero wavenumber by the fronts 
in $|A|$ and $|B|$, i.e. the sources, combined with the competition of 
counter-propagating waves of different wavenumbers \cite{RiArun}. A simple calculation
shows that right-traveling waves with a wavenumber $p_0=p_0^{(r)}$ are stable with respect to 
left-traveling waves with wavenumber $p_0=0$ only for
\bea
(p_0^{(r)})^2<\mu \, \frac{h_r-c_r}{h_r}.\label{e:counter}
\eea
Beyond this stability limit infinitesimal left-traveling waves are not suppressed by 
the right-traveling waves and grow. If the instability
is only convective the left-traveling waves are swept away and,
depending on the size of the domain 
of right-traveling waves, either perturb the next source or they grow to full 
amplitude forming a new domain of left-traveling waves. These dynamics ensue if the 
wavenumber selected by the source \cite{Ma94} 
is beyond the stability limit given by (\ref{e:counter}). We have numerically determined
the selected wavenumber. As shown in fig.\ref{f:quell}, it reaches the stability
boundary (\ref{e:counter}) at $h_r=1.3$. Thus, 
only for $h_r \ge 1.3$ 
the wavenumber selected by a source is stable to all counter-propagating waves and
the domains of left- and 
right-traveling waves become stationary and stable. These dynamics have also been 
discussed recently in the context of an investigation of sources \cite{HeSt99}.

\section{Conclusion}

In conclusion, we have performed a detailed  stability analysis of
standing waves with small group velocity and found  behavior that is quite different 
from that 
obtained for traveling waves. Due to the group velocity, the analogue of the Eckhaus parabola
is shifted with respect to the neutral curve and can be convex from below
and convex from above. The unstable solutions can be outside or inside
the parabola. In addition, a conservative long-wave instability can arise already at
lower order and generically a short-wave instability arises. The Newell-type
criterion for the instability of the band-center can be satisfied even for 
weak dispersion,
but it does not indicate the loss of stability of standing waves at all wave numbers. 

For group velocities of O(1) the equations discussed here can be viewed as model equations
that are obtained by reconstitution of the leading orders in the amplitude expansion.
As has been shown previously \cite{KnDe90,Kn92,MaVo92,Ve93},
in the asymptotic limit of small amplitudes non-local equations are obtained.
Within the non-local equations the stability properties are considerably simpler.
Our detailed analysis of the local (model) equations confirms that the non-local equations
are valid immediately above threshold. However, in particular for large systems and for parameters close to the Newell criterion the control-parameter 
range over which the non-local equations are valid can be extremely small.
Thus, already very close to threshold a cross-over can occur to the rich behavior discussed
here. A very interesting question is therefore whether this behavior is 
observable in experimental systems, as might be expected if the cross-over occurs very close to threshold.
In nematic liquid crystals the group velocities turn out to be 
quite small \cite{TrKr95,TrKr97}. This sytem is therefore 
a good candidate to investigate this question.

The stability analysis of standing waves gives sufficient conditions for the instability of 
traveling rectangles in two dimensions. It is in qualitative
agreement with experimental results on spatio-temporally chaotic waves
in nematic liquid crystals \cite{DeCa95,DeAh96,DeCa98}. Based on the recent 
calculation of the nonlinear coefficients in the Ginzburg-Landau equations \cite{TrKr97}, 
quantitative comparisons will be possible 
\cite{TrRi98}. 

In view of the richness of the presented results experiments on rotating
convection at small Prandtl numbers are of great interest, since in this system a Hopf bifurcation to standing waves has been predicted \cite{ClKn93}.

Finally we wish to emphasise that the phenomena treated in this work pertain to spatial 
modulations (in particular long-wave ones) of the two underlying traveling wave modes. 
The amplitude degeneracy occurring at $h_r = c_r$ in principle leads to other effects and 
necessitates the inclusion of higher-order terms \cite{CrKn88a,GoRo87}.

Financial support by DFG (Kr690/12) and by DOE (DE-FG02-92ER14303) is gratefully acknowledged.




\begin{thebibliography}{10}
\bibitem{scale}
One can rescale the amplitudes in order to have $c_r=1$, as is often done in the literature.
Then one has to replace in all formulas $c_i$ and $h$ by $c_i/c_r$ and $h/c_r$, respectively.
\end{thebibliography}

\end{document}